\documentstyle[12pt]{article}
\titlepage

\title{The Newtonian Potential in Gravitational Cohomology}
\author{  Richard Atkins \\
        richard.atkins@twu.ca \\
        Department of Mathematics\\ 
		Trinity Western University \\
		7600 Glover Road \\
		Langley, BC, V2Y 1Y1 Canada}
\date{}
\newtheorem{fact}{Fact}

\newtheorem{theorem}[fact]{Theorem}

\begin{document}
\maketitle
\begin{abstract}
We examine the Newtonian potential in gravitational cohomology. This is given by
a symmetric, two-index tensor field, which satisfies the wave equation in empty space.
Furthermore, the associated gravitational field
strength, obtained by applying the coboundary operator to the potential, 
is constructed for the case of a stationary point mass and shown to give the classical result.
\end{abstract} 
\newpage

\section{Introduction}
The quantization of electrodynamics was initiated in the 1920s through the efforts of Dirac, 
Heisenberg and Pauli. Fundamental to their approach was the description 
of the classical theory in terms of potentials, or 1-forms, whose coboundaries in the 
de Rham complex give the 2-form electromagnetic fields. The significance of this 
cohomological picture lies in the fact that it is the potentials that are to be 
quantized rather than the field strengths in the associated quantum field theory. 

The dynamics of a free particle under the influence of a
gravitational field is described by
\[ \ddot{x}^{\mu} = -\Gamma^{\mu}_{\nu \lambda}\dot{x}^{\nu}\dot{x}^{\lambda} \]
where $\Gamma^{\mu}_{\nu \lambda}$ is the Levi-Civita connection 
of the Lorentzian spacetime metric.
Comparing these geodesic equations to the equations of motion 
\[ \ddot{x}^{\mu} = -kF^{\mu}\hspace{0.01in}_{\nu} \dot{x}^{\nu} \]
of a charged particle in an electromagnetic field 
we see that the connection plays the role of the gravitational field strength. 
The question then arises whether a quantum 
formulation of gravitation also demands a characterization of the classical theory in terms of 
potentials, whose coboundaries give the connection.
More generally, it may be asked whether spacetime can be represented 
within some cohomological framework, apart from concerns of quantization. 

In \cite{aa}, we develop a cochain complex associated to a manifold $M$ endowed with a 
flat metric, whose cohomology is identified with the  \v{C}ech cohomology
of the sheaf of affine sections of $M$ (cf. \cite{bb}, \cite{kk}, \cite{mm} and
\cite{nn}). Moreover, it is shown that the Majumdar-Papapetrou spacetimes 
(cf. \cite{cc}, \cite{ee} and \cite{ff}) 
as well as the Schwarzschild black hole may be expressed in terms of potentials.
One may also envisage the inclusion of a curved metric $g_{\mu \nu}$ so that the 
relevent fields are the flat background metric $\eta_{\mu\nu}$, the spacetime metric
$g_{\mu\nu}$ and the gravitational potential $A=A_{\mu \nu}$. The field equations  
in empty space encompassing these three are 
\[ R_{\mu \nu}(A) -\frac{1}{2}R(A)g_{\mu \nu}=0 \]
where $R_{\mu \nu}(A)$ is the Ricci curvature of the connection 
$\Gamma^{\mu}_{\nu \lambda}(A)$,
related to the potential $A$, and $R(A)$ is the scalar curvature obtained by contraction 
with $g$: $R(A):= R_{\mu \nu}(A)g^{\mu \nu}$. The theory thus takes the shape of a  
modification of the bi-metric approach to gravitation that enjoyed attention for many
years (cf. \cite{ii}). It was in its mathematical simplicity that the appeal of bi-metric
geometry lay but the addition of a flat structure appeared to tarnish the splendor of
conventional general relativity and eventually it fell out of favour. 
However, as quantization of 
gravity remains one of the foremost outstanding problems in the natural sciences we reconsider
bi-metric models, now within the context of cohomology. 

In order to provide some justification for the course taken in \cite{aa} it is necessary
to study some of the elementary properties of the complex and its correlation to reality.
Accordingly, in this paper we look at the Newtonian potential $A_{\mu \nu}$ 
and derive  some of its features. First, it is shown that $A_{\mu \nu}$  satisfies 
the wave equation in empty space. Then we consider the field strength defined by the 
Newtonian potential. It portrays a gravitational  
influence propagating at the speed of light and which reduces to the usual Newtonian 
description in the stationary limit, as required.  This, in effect, reformulates the 
Newtonian theory in a manner that adopts general covariance and eschews action at a distance.

In the following section we review the definition and basic theorems of the gravitational 
cohomology presented in \cite{aa} in order to make the material herein self-contained. 
The cochain complex is built from symmetric-free $(n+2)$-tensor fields
\[ S_{\mu_{1}\cdots \mu_{n}\nu_{1}\nu_{2}} \]
which are skew-symmetric in the $\mu$ indices and symmetric in the $\nu$ indices. 
The gravitational potentials are representatives of such  tensors for  which $n=0$ and the
symmetric-free part of the field strengths are cocycles with $n=1$. 
In Section 3 some elementary formulas are derived for the calculus of fields corresponding to
a point source. The last section deals with the Newtonian potential.
It satisfies physically reasonable properties and 
the associated field strength obtained by applying the coboundary operator of the cochain 
complex to the potential results in the Newtonian force field expressed in Lorentz 
covariant form.

\section{Gravitational Cohomology}
This section briefly reviews some of the developments in \cite{aa}, 
to which the reader is referred for further elucidation and detail. We construct a 
cochain complex $(G^{*}(M), d)$ for a flat manifold, whose coboundary operator 
$d^{(1)}$ will define
the gravitational field strength associated to a given potential. $d^{(1)}$ plays the
analogous role in gravitation to the exterior derivative, acting on 1-form potentials, in
electrodynamics.

Let $M$ be a manifold with a flat metric $\eta_{\mu \nu} =$ $diag(+1,-1,-1,-1)$ and let 
$\nabla$ denote the Levi-Civita connection of $\eta_{\mu \nu}$.
$\nabla$ operates on tensor fields $C=C_{\mu_{1}...\mu_{n}}$ as follows:
\[ (\nabla C)_{\mu_{1}\cdots\mu_{n+1}}dx^{\mu_{1}}\otimes \cdots \otimes dx^{\mu_{n+1}} = 
dx^{\mu_{1}} \otimes \nabla_{\frac{\partial}{\partial x^{\mu_{1}}}}(C) \]
which may be written more compactly as
\[ \nabla_{\mu_{1}} C_{\mu_{2}\cdots\mu_{n+1}} \]

We raise and lower indices on symmetric connections with the flat metric:
\[ \Gamma_{\mu \nu\lambda}:= \eta_{\mu \sigma} \Gamma^{\sigma}_{\nu\lambda} \]
$\Gamma_{\mu \nu\lambda}$ may be decomposed into a symmetric part $\Gamma_{(\mu \nu\lambda)}$
and a symmetric-free part $F_{\mu\nu\lambda}$:
\[ \Gamma_{\mu \nu\lambda} = \Gamma_{(\mu \nu\lambda)} + F_{\mu\nu\lambda} \]
It is easy to see that the symmetric-free part satisfies 
\[ \frac{1}{2}F_{(\mu \nu \lambda)} = F_{\mu\nu\lambda} + F_{\nu\lambda\mu} + 
   F_{\lambda\mu\nu} =0 \]
This motivates the following definition.
For $n\geq 2$, let $G^{n}(M)$ denote the $(n+1)$-tensor fields 
$S_{\mu_{1}...\mu_{n+1}}$ on $M$ 
which are \\
(1) symmetric in the two rightmost indices $\mu_{n}$ and $\mu_{n+1}$, \\
(2) skew-symmetric in the $n-1$ leftmost indices $\mu_{1},...,\mu_{n-1}$, and \\
(3) which satisfy the symmetric-free condition
\begin{eqnarray*} \label{G}
S_{\mu_{1}\cdots\mu_{n+1}}  +  (-1)^{n}S_{\mu_{2}\cdots\mu_{n+1}\mu_{1}}  +  
S_{\mu_{3}\cdots\mu_{n+1}\mu_{1}\mu_{2}} +  
(-1)^{n}S_{\mu_{4}\cdots\mu_{n+1}\mu_{1}\mu_{2} \mu_{3}}+ \nonumber \\
S_{\mu_{5}\cdots\mu_{n+1}\mu_{1}\mu_{2} \mu_{3}\mu_{4}} +
\cdots  + (-1)^{n}S_{\mu_{n+1}\mu_{1}\cdots\mu_{n}} = 0 
\end{eqnarray*}
$G^{0}(M)$ shall designate the smooth functions on $M$ and $G^{1}(M)$, the symmetric tensor
fields $S_{\mu_{1}\mu_{2}}$ in two indices.
Skew-symmetrization of an $n$-tensor $C=C_{\mu_{1}...\mu_{n}}$  is defined by
\[  C_{[\mu_{1}\cdots\mu_{n}]} := 
   \frac{1}{n!}\sum_{\sigma \in S_{n}}sg(\sigma)C_{\mu_{\sigma(1)}\cdots\mu_{\sigma(n)}} \]
where $S_{n}$ is the set of permutations $\sigma$ on $n$ letters.
The exterior covariant derivative $d_{\nabla}$ is given by the formula
\[ (d_{\nabla}C)_{\mu_{1}\cdots\mu_{n+1}} = 
 \nabla_{[\mu_{1}}C_{\mu_{2}\cdots\mu_{n}]\mu_{n+1}} \]

For $n\geq 2$, define $\phi^{(n)}$ to be symmetrization of the two rightmost 
indices of an $(n+1)$-tensor:
\[ \phi^{(n)}(S)_{\mu_{1}\cdots\mu_{n-1}\nu \lambda} := S_{\mu_{1}\cdots\mu_{n-1}(\nu\lambda)}
   =\frac{1}{2}
(S_{\mu_{1}\cdots\mu_{n-1}\nu\lambda}+S_{\mu_{1}\cdots\mu_{n-1}\lambda\nu}) \]
and let $\psi^{(n)}$ be  skew-symmetrization of the $n$ leftmost  indices of an 
$(n+1)$-tensor, up to a constant multiple:
\[ \psi^{(n)}(S)_{\mu_{1}\cdots\mu_{n}\nu} := \frac{2n}{n+1}S_{[\mu_{1}\cdots\mu_{n}]\nu} \]
Define $d^{(0)}:G^{0}(M)\longrightarrow G^{1}(M)$ by $d^{(0)} = \nabla^{2}$ and
$d^{(n)}:G^{n}(M) \longrightarrow G^{n+1}(M)$ by $d^{(n)}= \phi^{(n+1)} \circ d_{\nabla}
  \circ \psi^{(n)}$, for $n\geq 1$. 
The  cochain complex $(G^{*}(M), d)$ is 
\[ 0 \longrightarrow G^{0}(M)  \stackrel{d^{(0)}}{\longrightarrow} G^{1}(M) 
    \stackrel{d^{(1)}}\longrightarrow  G^{2}(M) 
    \stackrel{d^{(2)}}\longrightarrow 
    G^{3}(M) \stackrel{d^{(3)}}\longrightarrow \cdots   \]
Since $\nabla$ is flat, $d^{(n+1)}\circ d^{(n)} =0$.

For our purposes we are interested only in the map 
$d^{(1)}:G^{1}(M) \longrightarrow G^{2}(M)$,
which defines the coboundary operator on gravitational potentials 
$A_{\mu \nu}\in G^{1}(M)$. 
This is given explicitly by
\begin{eqnarray} \label{fs}
d^{(1)}A_{\mu \nu \lambda} = \frac{1}{2}\nabla_{\mu}A_{\nu \lambda}
 -\frac{1}{4}\left( \nabla_{\nu}A_{\mu \lambda} + \nabla_{\lambda}A_{\mu \nu} \right) 
\end{eqnarray}
and produces the symmetric-free part $F_{\mu \nu\lambda}$ of the lowered Christoffel
symbol $\Gamma_{\mu \nu \lambda}$ of the connection.

The coboundary operators $d^{(n)}$ defined above 
may seem to be somewhat arbitrarily contrived. 
One partial validation of their form is that the symmetric-free component 
$F_{\mu \nu\lambda}$ of the 
lowered Christoffel symbols of the Levi-Civita connection of MP spacetimes as well as
the Schwarzschild black hole may be represented as the coboundary 
$F_{\mu \nu\lambda}=d^{(1)}A_{\mu \nu \lambda}$
of a gravitational potential $A_{\mu \nu}$. 
That $(G^{*}(M), d)$ is a natural cochain complex
associated to a flat manifold is justified, at least from a mathematical perspective, 
by the fact that the cohomology of the complex is isomorphic to the \v{C}ech cohomology
of the sheaf {\it Aff}, of affine sections of $M$; the group of sections 
{\it Aff}$(U)$  over an open subset $U$ of $M$ is defined
to be the kernel of the map $\nabla^{2}:G^{0}(U) \longrightarrow G^{1}(U)$.
This paper seeks to establish further support for the complex $(G^{*}(M), d)$ in describing
gravitation by investigating the behaviour of the Newtonian potential under
$d^{(1)}$; if the complex expresses a feasible approach to gravity then we expect
the coboundary (\ref{fs}) of the Newtonian potential to reduce to the conventional 
Newtonian force under stationary conditions. The remainder is devoted to this exploration.

\section{Elementary Formulas}

In this section we consider the fields related to a particle $Q$ moving through a flat 
background described by a four-manifold $M$ with a metric $\eta_{\mu \nu} = diag(+1,-1,-1,-1)$
and associated Levi-Civita connection $\nabla$. 
We shall work in flat coordinates $x=(t,{\bf x})$ on $M$ with respect to which covariant
differentiation becomes partial differentiation. 

It shall be supposed that the gravitational field propagates at the speed of light, 
which shall be set to unity by means of a suitable choice of units. 
The position of $Q$ as it moves through
$M$ is described by a 4-vector $q^{\mu}(t)= (t,{\bf q}(t))$. 
For an arbitrary point $x$ in spacetime we ask when and where would 
the emission of a gravitational influence from $Q$ be such that it would be felt at $x$. 
This leads to the definition of retarded time $t_{r}=t_{r}(x)$ defined by 
\begin{eqnarray} \label{ret}
t-t_{r}(x)=|{\bf x}-{\bf q}(t_{r}(x))|
\end{eqnarray}
Thus, a gravitational signal
emitted from $Q$ at position $(t_{r}(x),{\bf q}(t_{r}(x)))$ will, in the course of time,
propagate to $x$. Notice that $t_{r}(x)$ is a scalar field that assumes a value for 
each point $x$ in the manifold $M$.
The gravitational field of $Q$ will be constructed from the following covariant 
fields, which depend upon at most second-order derivatives of $q^{\mu}$: 
\begin{eqnarray}
C^{\mu}(x) & := & x^{\mu}-q^{\mu}(t_{r}(x)) \label{c} \\
U^{\mu}(x) & := & \frac{dq^{\mu}}{d\tau}|_{t=t_{r}(x)}
  \hspace{0.5in} \mbox {and} \label{u} \\
\dot{U}^{\mu}(x) & := & \frac{d^{2}q^{\mu}}{d \tau^{2}}|_{t=t_{r}(x)} \label{udot}
\end{eqnarray}
Here, $\tau$ is proper time as measured by $Q$. 
Let us look more closely at these fields and their physical interpretation.
$C^{\mu}(x)=(C^{0}(x),{\bf C}(x))$ is the difference between the point $x$ 
and the 4-position of $Q$, when its
gravitational influence, moving at the speed of light, would reach $x$. 
Clearly $C^{\mu}(x)$ is a lightlike vector for each $x \in M$:
\begin{eqnarray} \label{lightlike}
 C^{\mu}(x)C_{\mu}(x) = 0 
\end{eqnarray} 
and
\begin{eqnarray} \label{c0}
C^{0}(x) = t-t_{r}(x) = |{\bf x} - {\bf q}(t_{r}(x)) |
\end{eqnarray}
$C^{0}(x)$ measures
the time it takes for a light signal emitted from $Q$ at $(t_{r}(x),{\bf q}(t_{r}(x)))$
to reach $x$ and ${\bf C}(x)$ is the
displacement vector from ${\bf q}(t_{r}(x))$ to ${\bf x}$. If $Q$ were stationary
at position ${\bf q}$ in the $(t,{\bf x})$-coordinates then $C^{0}(x)=|{\bf C}(x)|$ 
would measure the distance $r$ between ${\bf x}$ and $Q$. 
$U^{\mu}(x)= (U^{0}(x),{\bf U}(x))$ is the 4-velocity of $Q$
at the moment a light signal from $Q$ would reach $x$. $U^{\mu}(x)$ is not the conventional 
4-velocity
of a particle, which is defined as a function of proper time or some other parameter,
but is in fact a vector {\it field}, defined for each $x\in M$. $\dot{U}^{\mu}(x)$
is the 4-acceleration of $Q$ at the moment a signal from $Q$ travelling at lightspeed 
would reach $x$; again, this is a field and not the usual function of a single variable.

By forming contractions of $C^{\mu}(x),U^{\mu}(x)$ and $\dot{U}^{\mu}(x)$ with the metric
$\eta_{\mu \nu}$ we may obtain scalar fields, the most important of which are: 
\begin{eqnarray}
X(x) & := & C^{\mu}(x) U_{\mu}(x) \hspace{0.5in} \mbox{and}    \label{x} \\
Y(x) & := & C^{\mu}(x) \dot{U}_{\mu}(x)
\end{eqnarray}
For $Q$ stationary at ${\bf q}$, $U_{\mu} = \eta_{0\mu}$ and $X$ becomes the distance $r$ to
the source $Q$:
\begin{eqnarray} \label{dist}
X(x)=C^{0}(x)= |{\bf x} - {\bf q}| = r
\end{eqnarray}
The fields $t_{r}, C^{\mu},U^{\mu},\dot{U}^{\mu}$ and $X$ are all defined on $M$
so they may be differentiated with respect to $x^{\mu}$. Next, we seek formulas for
these derivatives. We begin with $t_{r}(x)$, from which the partial derivatives 
of the others will follow.

It will be convenient to use  the relativistic scale factor $\gamma$:  
\begin{eqnarray} \label{gamma}
 \gamma (x)  :=  U^{0}(x) = \frac{dt}{d\tau}|_{t=t_{r}(x)} = 
   [1-(\frac{d{\bf q}}{dt}(t_{r}(x)))^{2}]^{-1/2} 
\end{eqnarray}
Observe that
\begin{eqnarray} \label{timeder}
\frac{d{\bf q}}{dt}(t_{r}(x)) = \frac{d{\bf q}}{d\tau}|_{t=t_{r}(x)}
 \frac{d\tau}{dt}(t_{r}(x)) = {\bf U}(x)\gamma^{-1}(x)
\end{eqnarray}
Differentiating (\ref{ret}) implicitly with respect to the time coordinate $x^{0}=t$ gives:
\[ 1-\frac{\partial t_{r}}{\partial t} = - 
   \frac{{\bf x}-{\bf q}(t_{r})}{|{\bf x}-{\bf q}(t_{r})|} \cdot \frac{d{\bf q}}{dt}(t_{r}) 
   \frac{\partial t_{r}}{\partial t} \]
Hence,
\[ \frac{\partial t_{r}}{\partial t} = 
  \frac{|{\bf x}-{\bf q}(t_{r})|}{|{\bf x}-{\bf q}(t_{r})| - 
  ({\bf x}-{\bf q}(t_{r}))\cdot \frac{d{\bf q}}{dt}(t_{r}) }   \]
By (\ref{c}), (\ref{c0}), (\ref{gamma}) and (\ref{timeder}),
\[ \frac{\partial t_{r}}{\partial t} = \frac{C^{0}}{C^{0}-{\bf C}\cdot{\bf U} \gamma^{-1}}
   = \frac{C^{0}}{(C^{0}U^{0}-{\bf C}\cdot{\bf U} )\gamma^{-1}} \]
Definition  (\ref{x}) allows us to rewrite this expression more compactly:
\begin{eqnarray} \label{par1}
\frac{\partial t_{r}}{\partial t} = \gamma X^{-1}C_{0}
\end{eqnarray}  
where we have lowered the index on $C$.   

Now consider the implicit partial derivative of (\ref{ret}) with respect to a 
spatial coordinate $x^{i}$, $1\leq i \leq 3$. This yields
\[ -\frac{\partial t_{r}}{\partial x^{i}} = 
   \frac{{\bf x}-{\bf q}(t_{r})}{|{\bf x}-{\bf q}(t_{r})|} \cdot 
   ( {\bf e}_{(i)}- \frac{d{\bf q}}{dt}(t_{r})  \frac{\partial t_{r}}{\partial x^{i}} ) \]
where ${\bf e}_{(i)}$ denotes the $i^{th}$ standard basis vector in $\Re^{3}$.
Solving algebraically we obtain
\[ \frac{\partial t_{r}}{\partial x^{i}} = -
   \frac{x^{i}-q^{i}(t_{r})}{|{\bf x}-{\bf q}(t_{r})| - 
  ({\bf x}-{\bf q}(t_{r}))\cdot \frac{d{\bf q}}{dt}(t_{r}) }  \]
Applying (\ref{c}), (\ref{c0}), (\ref{gamma}) and (\ref{timeder}) once again,
this may be written 
\[  \frac{\partial t_{r}}{\partial x^{i}} 
   = -\frac{C^{i}}{(C^{0}U^{0}-{\bf C}\cdot{\bf U} )\gamma^{-1}}  \]
Lowering the index on $C$ gives
\begin{eqnarray}\label{par2}
\frac{\partial t_{r}}{\partial x^{i}} = \gamma X^{-1}C_{i}
\end{eqnarray}
(\ref{par1}) and (\ref{par2}) may be combined into a single equation:
\begin{eqnarray} \label{part}
t_{r},_{\mu} = \gamma X^{-1}C_{\mu}
\end{eqnarray}

Calculating the partial derivatives of the fields (\ref{c})-(\ref{udot}) and
(\ref{x}) is now simply a matter of invoking the definitions along with 
equation (\ref{part}).
For instance,
\begin{eqnarray*}
\frac{\partial C^{\mu}(x)}{\partial x^{\nu}} & = & 
\frac{\partial x^{\mu}}{\partial x^{\nu}}-\frac{\partial q^{\mu}(t_{r}(x))}{\partial x^{\nu}} 
 \hspace{0.62in} \mbox{by (\ref{c})} \\
& = & \delta^{\mu}_{\nu} - \frac{dq^{\mu}}{dt}(t_{r}) \frac{\partial t_{r}}{\partial x^{\nu}} \\
& = & \delta^{\mu}_{\nu} - \frac{dq^{\mu}}{d\tau}|_{t=t_{r}(x)}\frac{d\tau}{dt}(t_{r})
       \frac{\partial t_{r}}{\partial x^{\nu}} \\
& = & \delta^{\mu}_{\nu} - U^{\mu}\gamma^{-1}\gamma X^{-1}C_{\nu}  
      \hspace{0.5in} \mbox{by (\ref{u}), (\ref{gamma}) and (\ref{part})} \\
& = &   \delta^{\mu}_{\nu} - X^{-1}U^{\mu}C_{\nu}  
\end{eqnarray*}
The other partial derivatives may be handled in a similarly straightforward, 
albeit tedious, manner and not wishing to weary the reader we content ourselves
by merely stating the relevent results in the following theorem.
\begin{theorem} \label{derivatives} \hspace{0in} \\
$C_{\mu},_{\nu}=\eta_{\mu \nu} - X^{-1}U_{\mu}C_{\nu}$  \\
$U_{\mu},_{\nu}=X^{-1}\dot{U}_{\mu}C_{\nu}$  \\
$\dot{U}_{\mu},_{\nu}=X^{-1}\ddot{U}_{\mu}C_{\nu}$ \hspace{0.2in} where \hspace{0.1in} 
$\ddot{U}_{\mu}(x)= \frac{d^{3}q_{\mu}}{d\tau^{3}}|_{t=t_{r}(x)}$  \\
$X,_{\nu}=U_{\nu}+hC_{\nu}$ \hspace{0.27in} where \hspace{0.1in}
 $h=X^{-1}(Y-1)$  \\
$\Box U_{\mu} = 2X^{-1}\dot{U}_{\mu}$  
\end{theorem}

\section{The Newtonian Potential}

The potential in Newtonian gravity is a scalar field; 
we must find its counterpart $A_{\mu \nu}$ in $G^{1}(M)$. 
It is expected to be 
constructed from covariant fields related to a point source $Q$, which depend 
upon derivatives of at most first order in the position of $Q$, as is the case for
the Li\'{e}nard-Wiechert potentials in electrodynamics. This leaves us with $C^{\mu},U^{\mu}$
and $X$ as the basic building blocks for $A_{\mu \nu}$. 
Furthermore, when $Q$ is stationary the expression
for $A_{\mu \nu}$ must in some regard reflect the Newtonian form 
\begin{eqnarray} \label{Newt}
 V=\frac{c}{r} 
\end{eqnarray}
where $c$ is a constant and $r$ is distance from the source $Q$. 
Without belabouring these heuristic arguments we suggest:
\begin{eqnarray}
A_{\mu \nu} := aX^{-1}U_{\mu}U_{\nu} 
\end{eqnarray}
where the constant $a$ shall be determined later. For stationary $Q$, $U_{\mu}$ reduces to
$U_{\mu} = \eta_{0 \mu}$ and $X$ becomes $r$, as observed in (\ref{dist}). 
So in this case the only non-zero component
of $A_{\mu \nu}$ would be 
\[ A_{00} = \frac{a}{r} \]
in accordance with (\ref{Newt}).

Observe that $X^{-1}U_{\mu}$ is the Li\'{e}nard-Wiechert potential up to a constant multiple.
In empty space it satisfies the wave equation
\begin{eqnarray} \label{lw}
\Box X^{-1}U_{\mu} = 0 
\end{eqnarray}
Consider now the d'Alembertian operator acting on $X^{-1}U_{\mu}U_{\nu}$:
\[ \begin{array}{lll}
\Box X^{-1}U_{\mu}U_{\nu} & = & (\Box X^{-1}U)_{\mu}U_{\nu} + 2 (X^{-1}U_{\mu})^{,\sigma}
  U_{\nu},_{\sigma} + X^{-1}U_{\mu}(\Box U)_{\nu} \\
& = & 2[-X^{-2}(U^{\sigma}+hC^{\sigma})U_{\mu}+X^{-2}\dot{U}_{\mu}C^{\sigma}]
      X^{-1}\dot{U}_{\nu}C_{\sigma} + \\ 
& &   X^{-1}U_{\mu}2X^{-1}\dot{U}_{\nu} \\
& = & -2X^{-2}U_{\mu}\dot{U}_{\nu} +2X^{-2}U_{\mu}\dot{U}_{\nu} \\
& = & 0
\end{array} \]
In this derivation we have used formulas from Theorem \ref{derivatives} as well as 
equations (\ref{lightlike}), (\ref{x}) and (\ref{lw}). To summarize:
\begin{theorem}
The Newtonian potential $A_{\mu \nu}$ satisfies the wave equation 
\[ \Box A_{\mu \nu} =0\]
in empty space.
\end{theorem}

Our final concern is the calculation of the gravitational field tensor derived from
$A_{\mu \nu}$. First consider the covariant derivative of the Newtonian potential
\[ \begin{array}{lll}
\partial_{\mu}A_{\nu\lambda} & = & a(X^{-1}U_{\nu}U_{\lambda}),_{\mu} \\
 & = & aX^{-2}[-(U_{\mu}+hC_{\mu})U_{\nu}U_{\lambda} + C_{\mu}\dot{U}_{\nu}U_{\lambda} +
       C_{\mu}U_{\nu}\dot{U}_{\lambda}]
\end{array} \]
Suppose henceforth that 
$Q$ is stationary at the origin of the flat coordinate system: ${\bf q} =(0,0,0)$.
Then 
\[ 
U_{\mu} =  \eta_{0\mu};  \hspace{0.2in}
\dot{U}_{\mu}  = 0; \hspace{0.2in}
X  = C_{0} =  r; \hspace{0.2in} C_{i} = -x^{i}; 
\hspace{0.2in}h  =  -r^{-1}  \]
By substituting these into the above expression for  $\partial_{\mu}A_{\nu\lambda} $
we obtain
\begin{eqnarray} \label{A}
 \partial_{\mu}A_{\nu\lambda}  =  \left\{ 
\begin{array}{cl}
-\frac{ax^{\mu}}{r^{3}} & \hspace{0.3in} \mbox{if $\nu=\lambda=0$ and $\mu\neq 0$} \\
0 & \hspace{0.3in} \mbox{otherwise}
\end{array}  \right. 
\end{eqnarray}
Recall that the coboundary of the Newtonian potential is
defined by (\ref{fs}):
\begin{eqnarray} \label{F}
F_{\mu \nu \lambda}:= d^{(1)}A_{\mu \nu \lambda} = \frac{1}{2}\partial_{\mu}A_{\nu \lambda}
 -\frac{1}{4}\left( \partial_{\nu}A_{\mu \lambda} + \partial_{\lambda}A_{\mu \nu} \right)  
\end{eqnarray} 
Consider a particle $P$ of sufficiently small mass that it does not affect the gravitational
field produced by $Q$ in any substantial way. Its 4-position will be denoted
$x^{\mu}(t) = (t,{\bf x}(t))$  and $s$ shall designate its proper time. 
Assuming that the lowered Christoffel symbol has a vanishing symmetric part,
\[ \Gamma^{i}_{\nu \lambda} = F^{i}\hspace{0.01in}_{\nu \lambda} = 
   -F_{i \nu \lambda} \]
for $1\leq i \leq 3$.
The equations of motion governing
$P$ are  then
\[ \frac{d^{2}x^{i}}{ds^{2}} = -\Gamma^{i}_{\nu \lambda} 
  \frac{dx^{\nu}}{ds}\frac{dx^{\lambda}}{ds} = 
  F_{i \nu \lambda}   \frac{dx^{\nu}}{ds}\frac{dx^{\lambda}}{ds}
 \]
Placing the values of $\partial_{\mu}A_{\nu\lambda}$ from (\ref{A}) into (\ref{F})
results in the equations:
\[
\frac{d^{2}x^{i}}{ds^{2}} = 
\frac{1}{2}\partial_{i} A_{\nu\lambda}\frac{dx^{\nu}}{ds}\frac{dx^{\lambda}}{ds} =
     -\frac{ax^{i}}{2r^{3}}  \frac{dx^{0}}{ds}\frac{dx^{0}}{ds} 
\]
which for velocities of $P$ small compared to the speed of light becomes
\[ \frac{d^{2}x^{i}}{dt^{2}} = -\frac{ax^{i}}{2r^{3}}  \]
These are, in fact, the equations of motion for a mass in a Newtonian gravitational field
if 
\[ a = 2GM \]
where $G$ is Newton's constant and $M$ is the mass of $Q$. This is the desired
outcome.
\begin{theorem}
The potential  \[ A_{\mu\nu} = 2GMX^{-1}U_{\mu}U_{\nu} \]
yields the Newtonian gravitational theory for a stationary particle of
mass $M$.
\end{theorem}

We have required that the symmetric part $\Gamma_{(\mu \nu\lambda)}$ of the connection 
be zero; a non-vanishing symmetric part would amount to a modification of the 
Newtonian force law.

\end{document}